\documentclass[%
reprint,
superscriptaddress,
nofootinbib,
 amsmath,amssymb,
 aps,
 prl,
altaffilsymbol
]{revtex4-2}

\usepackage{graphicx}
\usepackage{dcolumn}
\usepackage{bm}

\usepackage[utf8]{inputenc}
\usepackage{ulem,color}

\newcommand{\dlt}[1]{}
\newcommand{\del}[1]{}

\newcommand{\ccs}[1]{}

\newcommand{\vo}{VO$_2$}

\newcommand{\vowtie}{VOwtie}
\newcommand{\diavolo}{diaVOlo}

\begin{document}



\title{Reconfigurable plasmonic hot spots enabled by composite \vo{}--gold plasmonic antennas}


\author{Rostislav Řepa}
\email{rostislav.repa@vutbr.cz}
\author{Jiří Kabát}
\author{Tomáš Šikola}
\author{Vlastimil Křápek}
\email{krapek@vutbr.cz}
\affiliation{Brno University of Technology, Central European Institute of Technology, Purkyňova 123, 612 00 Brno, Czechia}
\affiliation{Brno University of Technology, Institute of Physical Engineering, Technická 2, 616 69 Brno, Czechia}

\date{\today}

\begin{abstract}
We theoretically investigate the formation of electric and magnetic hot spots with reconfigurable plasmonic antennas. We consider three material systems offering different levels of reconfigurability: gold with the static response, vanadium dioxide which allows for ON/OFF switching, and composite gold--vanadium dioxide material platform which offers a possibility to switch between the electric and magnetic hot spot within a single antenna. Using bowtie and diabolo antennas as a case study, we evaluate optical response functions (scattering and absorption cross-sections, electric and magnetic field enhancement). We demonstrate that the composite material system brings, in addition to enhanced reconfigurability, also novel features of plasmonic antennas, such as strong optical absorption and a joint electric-magnetic hotspot.
\end{abstract}


\maketitle

{\it Introduction} 
Plasmonic antennas~\cite{Novotny2011} (PAs) enable nanoscale control of light by means of the excitation of localized surface plasmons (LSPs). One of their most desired functionality is the formation of hot spots: deeply subwavelength volumes in which the electromagnetic field supported by PAs is particularly strong. When an object is inserted into the hot spot of a PA, its interaction with the light is significantly enhanced. Enhancement factors for the electric component of the field as large as 10$^2$--10$^3$ are being commonly reported~\cite{Kneipp2013,PhysRevB.71.235420,Chikkaraddy2016}. Consequently, hot spots are exploited in numerous applications that rely on strong light-matter interactions. They include sensing~\cite{doi:10.1021/acs.accounts.1c00682}, catalysis~\cite{doi:10.1021/acs.nanolett.3c00219,doi:10.1021/acs.accounts.9b00234}, surface enhanced Raman spectroscopy~\cite{C4CP04946B}, plasmon enhanced photoluminescence~\cite{Kinkhabwala2009}, plasmon enhanced electron paramagnetic resonance~\cite{https://doi.org/10.1002/smtd.202100376}, strong plasmon-exciton coupling~\cite{Bitton2020}, optical trapping~\cite{PEI2026131293}, or local heat management~\cite{doi:10.1021/acs.jpcc.6b03644}.

\begin{figure}
\includegraphics[clip,width=\columnwidth]{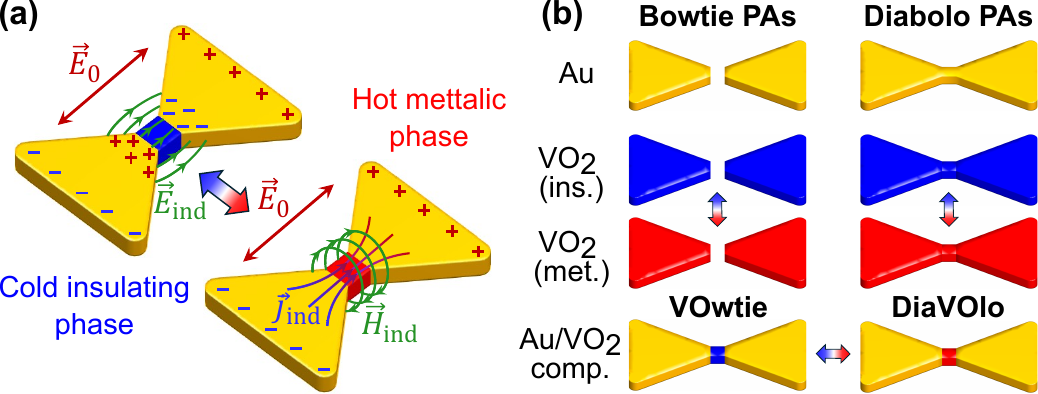}
\caption{\label{fig1}
(a) Bowtie and diabolo antennas with schemes of their operation and formation of hot spots. (b) Bowtie (left) and diabolo (right) antennas involved in our study: gold (first row), vanadium dioxide (second and third row), and composite (fourth row). The color of individual PA parts indicates their material composition: gold (yellow), metallic \vo{} (red), and insulating \vo{} (blue).
}

\end{figure}

The most common PA designs supporting hot spots are a gap between two parts of a dimer PA~\cite{PhysRevB.71.235420} or a sharp tip or edge of a monolithic PA~\cite{doi:10.1126/science.aah5243}, with advanced concepts including the exploitation of Babinet's principle in the dimer plasmonic apertures~\cite{doi:10.1021/nl103817f,PhysRevApplied.13.054045} or hierarchical plasmonic nanostructures~\cite{https://doi.org/10.1002/smll.202205659} also being developed. It is noteworthy that the enhancement of the electric and magnetic fields is often not equally strong. Based on the dominant component, the hot spots can be classified as electric (most frequent, with the majority of the previous references related to them), magnetic (also frequent~\cite{https://doi.org/10.1002/smtd.202100376,doi:10.1021/nl103817f,CalandriniCereaDeAngelisZaccariaToma+2019+45+62}), or mixed (those with a balanced enhancement of both fields are rather rare~\cite{doi:10.1021/acsphotonics.6b00857,krapek_independent}). Bowtie and diabolo PAs~\cite{Zhou:09,doi:10.1021/nl103817f} [Fig.~\ref{fig1}(a)] are particularly attractive for the formation of electric and magnetic hot spots, respectively. Their shape similarity (they both consist of triangular metallic wings which are coupled capacitively in the case of the bowtie and conductively in the case of the diabolo) has been recognized~\cite{krapek_independent,doi:10.1021/nl103817f} and led to a proposal of a reconfigurable PA which can be switched between the bowtie and the diabolo functionality~\cite{krapek_independent}.

PAs composed of conventional metals have their functionality fixed after their fabrication. Driven by the desire for reconfigurable and tunable PAs, exploitation of phase-change materials (PCMs) is being considered. Vanadium dioxide (\vo{}) represents one of the most popular PCMs for plasmonic applications. It exhibits a metal-insulator transition (MIT) with a convenient transition temperature around 68~$^\circ$C, which can be triggered by thermal, electrical, mechanical, or optical stimuli~\cite{LIU2018875}. During the MIT, the material exhibits significant changes in electrical conductivity (five orders of magnitude within a picosecond timescale) and optical properties (especially in the infrared region)~\cite{LIU2018875,D4NA00338A}. However, \vo{} also possesses some unfavourable properties, such as low optical conductivity even in the metallic phase and low Q factor of plasmon resonances~\cite{kabát2026tuninglocalizedsurfaceplasmons}, as well as restricted functionality, often limited to ON/OFF switching~\cite{10.1063/5.0028093}. \ccs{We need more external citations in this part}

In our paper, we inspect the possibility to realize reconfigurable plasmonic hot spots using \vo{}. To overcome its unfavourable properties, we propose to combine \vo{} as the active element and a conventional metal (gold) as a good conductor into composite PAs using a planar arrangement of both materials. Specifically, we focus on the bowtie and diabolo PAs with the gold wings and \vo{} bridge [Fig.~\ref{fig1}(a)], where switching between the bowtie and the diabolo functionality shall be achieved by switching \vo{} between the insulating and the conducting phases. To assess the performance of the proposed composite material platform, we also inspect homogeneous bowtie and diabolo PAs made of either gold or \vo{}. We demonstrate that the composite PAs support hot spots with unique properties (balanced electric and magnetic enhancement), exhibit extensive tunability (in terms of the resonant energy and scattering and absorption efficiencies, but not so much in terms of the electric and magnetic enhancement), and outperform pure \vo{} PAs in terms of the magnitude of the enhancements.

{\it Methods}
Theoretical modeling was performed using a formalism of classical electrodynamics with the numerical solution performed using the finite element method (software package Comsol Multiphysics). The geometry of the modeled PAs was set as follows: the length $L=500\ \mathrm{nm}$, the width $W=300\ \mathrm{nm}$, the length and the width of the bridge $B_L=40\ \mathrm{nm}$ and $B_W=40\ \mathrm{nm}$, respectively, the thickness $T=30\ \mathrm{nm}$. The dielectric function of gold was taken from Ref.~\cite{PhysRevB.86.235147} and the dielectric function of \vo{} was taken from Ref.~\cite{https://doi.org/10.1002/andp.201900188}. 

{\it Results} 
The bowtie and diabolo PAs are schematically shown in Fig.~\ref{fig1}(a). They consist of two triangular metallic wings with the proximal vertices either separated by an insulating gap (bowtie) or connected with a conducting bridge (diabolo). When excited with a plane wave polarized parallel to the long axis of the PA [as indicated by red arrows in Fig.~\ref{fig1}(a)], the LSP resonances result in the formation of hot spots, as indicated in Fig.~1(a). In the case of the bowtie PA, the charge is concentrated on the sides of the insulating gap, resulting in the electric hot spot. In the case of the diabolo PA, the current is funnelled through the conducting bridge, resulting in the magnetic hot spot. The properties of the bowtie and diabolo PAs, their LSP resonances, and hot spots have been thoroughly studied~\cite{Zhou:09,doi:10.1021/nl103817f,krapek_independent,PhysRevApplied.13.054045,doi:10.1021/jz302018x,doi:10.1021/nl203811q}. 

Figure~\ref{fig1}(b) represents 8 PAs involved in our comparison. The first set consists of two gold PAs: the bowtie and the diabolo. These PAs are static; they cannot be switched. The second set consists of four \vo{} PAs: the metallic bowtie and diabolo, and the insulating bowtie and diabolo which do not support plasmon resonance. Here, switching is possible, but only between the operational (metallic) PA and the non-operational insulating structure (e.g., between the metallic \vo{} bowtie antenna supporting an electric hot spot and the insulating \vo{} bowtie structure which lacks any functionality of a PA). We refer to this switching between a certain functionality and no functionality as ON/OFF switching. Finally, the third set includes two composite PAs, the \vowtie{} and the \diavolo{}, which can be mutually reconfigured from one to the other (and vice versa) by switching the central \vo{} element from the insulating gap to the metallic bridge (and vice versa). In this way, a switching between the electric and the magnetic hot spot can be expected.

In the following, we compare optical properties of these 8 PAs. First, we inspect PAs of the same geometry separately (i.e., the set of 4 bowties and the set of 4 diabolos) to address the effect of the involved material system. Next, we focus on the switching, comparing \vowtie{} and \diavolo{}, and, as a reference, also gold bowtie and diabolo (although they can only be replaced by one another, not dynamically switched). PAs are illuminated with a plane wave polarized along the long axis of PAs, as shown in Fig.~\ref{fig1}(a).

\begin{figure}[ht!]
\includegraphics[clip,width=\columnwidth]{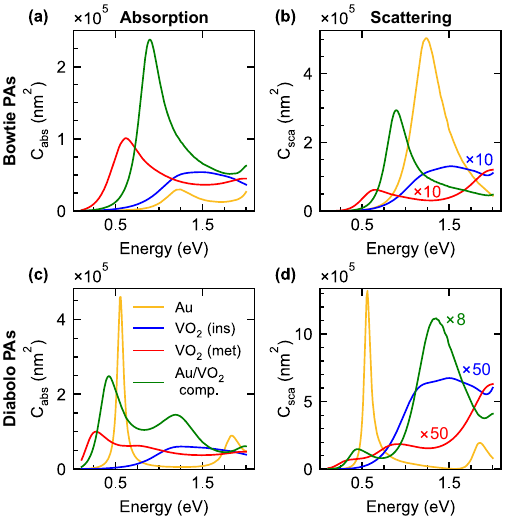}
\caption{\label{fig2}
Optical cross sections of bowtie and diabolo PAs. (a,b) Absorption cross section (a) and scattering cross section (b) of the bowtie PAs: gold bowtie PA (yellow line), metallic \vo{} bowtie PA (red line), insulating \vo{} bowtie structure (blue line), and composite \vowtie{} PA (green line). (c,d) Absorption cross section (c) and scattering cross section (d) of the diabolo PAs: gold diabolo PA (yellow line), metallic \vo{} diabolo PA (red line), insulating \vo{} diabolo structure (blue line), and composite \diavolo{} PA (green line). 
}
\end{figure}

To identify the spectral position of the hot spots and related LSP modes, we inspect optical cross sections (the absorption and the scattering cross section), defined as the power absorbed and scattered by the PA divided by the intensity of the illumination, respectively. Fig.~\ref{fig2} shows the spectral dependence of both the cross sections and Tabs.~\ref{tab1},~\ref{tab2} list several parameters relevant for the performance of PAs which were derived from the spectra of Fig.~\ref{fig2}.

\begin{table}[h!]
\caption{\label{tab1}
Optical parameters of the bowtie PAs: LDB energy defined as the central position of the absorption peak ($E$), peak values of the normalized absorption ($Q_\mathrm{abs}$) and scattering $Q_\mathrm{scat}$ cross section, and SAR ($Q_\mathrm{scat}/Q_\mathrm{abs}$). \ccs{Other candidates: Q factor, epsilon, and material Q factor.}
}
\begin{tabular}{lcccc}
\hline
\hline
& $E$~(eV) & $Q_\mathrm{abs}$ & $Q_\mathrm{scat}$ & SAR \\
\hline
gold bowtie & 1.25 & 0.381 & 6.36 & 16.7 \\
\vowtie{} & 0.890 & 3.02 & 3.73 & 1.24 \\
metallic \vo{} bowtie & 0.620 & 1.28 & 0.0770 & 0.0604 \\ 
\hline
\hline
\end{tabular}
\end{table}

\begin{table}[h!]
\caption{\label{tab2}
Optical parameters of the diabolo PAs: LDB energy defined as the central position of the absorption peak ($E$), peak values of the normalized absorption ($Q_\mathrm{abs}$) and scattering $Q_\mathrm{scat}$ cross section, and SAR ($Q_\mathrm{scat}/Q_\mathrm{abs}$). 
}
\begin{tabular}{lcccc}
\hline
\hline
& $E$~(eV) & $Q_\mathrm{abs}$ & $Q_\mathrm{scat}$ & SAR \\
\hline
gold diabolo & 0.560 & 5.84 & 16.8 & 2.87 \\
\diavolo{} & 0.420 & 3.16 & 0.229 & 0.0726 \\
metallic \vo{} diabolo & 0.280 & 1.27 & 0.0140 & 0.0112 \\
\hline
\hline
\end{tabular}

\end{table}

\begin{figure}[ht!]
\includegraphics[clip,width=\columnwidth]{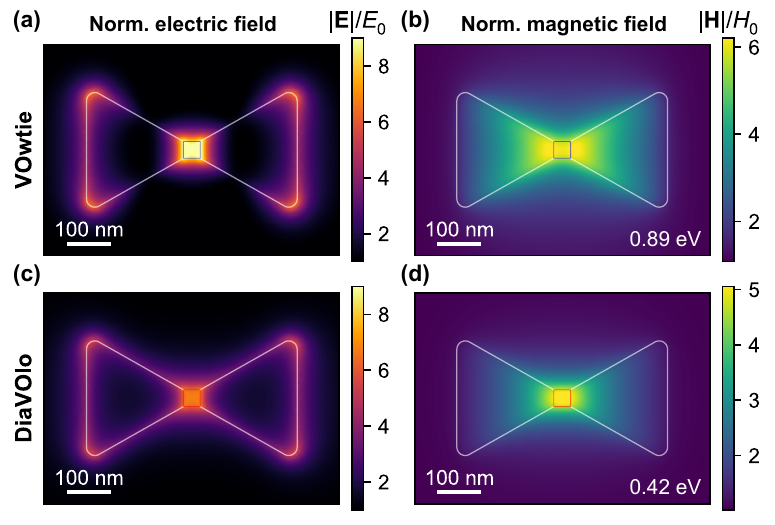}
\caption{\label{fig3}
Normalized electromagnetic field produced by the \vowtie{} and \diavolo{} PAs: planar cross sections at the height of 10~nm above the upper surface of PAs. PA boundaries are indicated by the white line. (a,b) The electric (a) and the magnetic (b) field of the \vowtie{} PA. (c,d) The electric (c) and the magnetic (d) field of the \diavolo{} PA.}
\end{figure}

All bowtie PAs (gold, metallic \vo{}, and \vowtie{}) exhibit a single peak in both cross sections in the spectral range between 0.6--1.3~eV. This peak is assigned to the longitudinal dipole bonding (LDB) mode (as defined in Ref.~\cite{krapek_independent}) responsible for the formation of the electric hot spot based on the analysis of induced fields, of which an example is shown in Fig.~\ref{fig3}. It is customary to define normalized cross sections as cross section divided by the PA geometric cross section (i.e., the area of PA, around $8\times 10^4$~nm$^3$ in our case), and peak values of the normalized cross sections are listed in Tab.~\ref{tab1} together with the scattering to absorption ratio (SAR) defined as the peak scattering cross section divided by peak absorption cross section. They demonstrate strong plasmonic response of gold bowtie and \vowtie{}, with the optical cross sections reading several multiples of the PA geometric cross section, and decent plasmonic response of the metallic \vo{} bowtie. We also readily observe the distinct character of individual bowties. The gold bowtie shows a strong scattering ($5.0\times 10^5$~nm$^2$) and decent absorption ($3.0\times 10^4$~nm$^2$). The \vowtie{} exhibits somewhat reduced scattering (reaching 59~\% of its gold-bowtie equivalent) but pronouncedly enhanced absorption ($7.9\times$ of its gold-bowtie equivalent\ccs{counterpart?}). Interestingly, the absorption and the scattering are nearly balanced for the \vowtie{}, evidenced by the near-one SAR.  We stress that the pronounced difference between the gold bowtie and the \vowtie{} is related to a minor material modification of the gap, which represents only about 5~\% of the total PA volume. However, this is not surprising, as the electric field exhibits a hot spot in the PA gap.

The metallic \vo{} bowtie is inferior to the gold bowtie and \vowtie{}, with considerably lower absorption cross section and negligible scattering cross section, naturally related to high optical absorption of \vo{}. At the LDB energy (1.25~eV in the gold bowtie and 0.62~eV in the metallic \vo{} bowtie), the dielectric function \vo{} reads $-10.13 + i12.12$ with the intrinsic Q factor of 0.84, while the dielectric function of gold reads $-40.30 + i1.85$ with considerably higher intrinsic Q factor of 21.8. Still, there are two findings of interest: (1) The normalized absorption cross section exceeding unity in the resonance highlights that the optical response is dominated by LSP. For comparison, a planar layer of the same material would yield a high reflection of 67~\% and only a moderate absorption of 17~\%. \ccs{multiple reflections neglected} (2) Particularly low SAR of 0.0604 makes the metallic \vo{} bowtie suitable for applications in the energy harvesting \ccs{only for heat, though} or antireflection coating. For completeness, we include to our comparison also the insulating \vo{} bowtie. This structure is not plasmonically active, and exhibits only a weak absorption response above 1.2~eV related to the material absorption in \vo{}.


All diabolo PAs (gold, metallic \vo{}, and \diavolo{}) exhibit the lowest-energy peak in both cross sections in the spectral range between 0.3--0.6~eV, attributed to the longitudinal dipole (LD) mode responsible for the formation of the magnetic hot spot (see Ref.~\cite{krapek_independent} for details). The assignment of the peak to the LD mode again relies on the analysis of the induced field, as exemplified in Fig.~\ref{fig3}. The role of the material composition is mostly similar to the bowties, with the exception of the \diavolo{} PA. The gold diabolo PA provides efficient scattering and absorption, the metallic \vo{} diabolo PA provides moderate scattering and very low SAR, and the insulating \vo{} diabolo structure lacks any plasmonic response. The \diavolo{} PA offers a sizeable absorption (although weaker than the gold diabolo PA), but its scattering is negligible (similar to metallic \vo{} PAs and unlike the \vowtie{}). We attribute this to the large optical losses in the metallic \vo{} and to the fact that the absorption in the diabolo bridge is significantly larger than the absorption in the bowtie gap due to a large current density in the bridge (while in the bowtie, the current is not transferred through the gap). The low SAR and the large absorption make the \diavolo{} PA well suited for the same applications as the metallic \vo{} bowtie and diabolo PAs (among others, the antireflection coatings), with better performance as its absorption is approximately twice as large.



\begin{figure}[ht!]
\includegraphics[clip,width=\columnwidth]{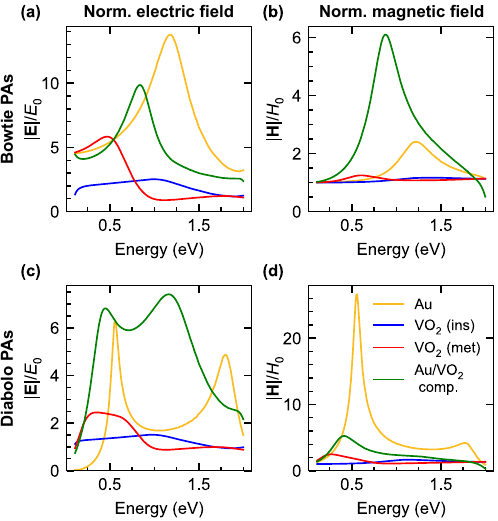}
\caption{\label{fig4}
Normalized electromagnetic field produced by the bowtie and diabolo PAs: Spectral dependence at the point above the center of the PA, 10~nm above the upper surface. (a,b) The electric (a) and the magnetic (b) field of the bowtie PAs. (c,d) The electric (c) and the magnetic (d) field of the diabolo PAs. The colors indicate the material composition: gold (yellow), metallic \vo{} (red), and composite (green).}
\end{figure}

In the following, we will discuss the electromagnetic field produced by the PAs (normalized by the incident plane wave) with a special focus on the hot spots. Fig.~\ref{fig3} shows the spatial distribution of the field for the \vowtie{} and \diavolo{} PAs at the energy of the LDB (0.89~eV) and LD (0.42~eV) mode, respectively (i.e, at the energy of the LSP resonance supporting the corresponding hot spot). We indeed observe a formation of the electric hot spot by the \vowtie{} PA [Fig.~\ref{fig3}(a)] and the magnetic hot spot by the \diavolo{} PA [Fig.~\ref{fig3}(d)] in the central part (the gap or the bridge) of PAs. Interestingly, we also observe rather unexpected localization of the remaining field components [i.e., the magnetic field of the \vowtie{} in Fig.~\ref{fig3}(b) and the electric field of the \diavolo{} in Fig.~\ref{fig3}(c)] in the central part of PAs. Consequently, the hot spots supported by the \vowtie{} and \diavolo{} PAs have strong mixed electric-magnetic character\, attributed to the non-zero conductivity of the gap in the \vowtie{} and the low conductivity (when compared to gold) of the bridge in the \diavolo{}. This feature is not observed in gold or metallic \vo{} PAs, as demonstrated in Fig.~\ref{fig4} and Tab.~\ref{tab3}.


For a quantitative comparison of all PAs, we inspect normalized fields at the hot spot, which we represent by the point horizontally in the center of the PA and vertically at the height of 10~nm above the upper surface of the PA, where all the fields related to the hot spot (i.e., the electric field of the bowties and the magnetic field of the diabolos) are near their maximum for all PAs. The spectral dependence of the fields is shown in Fig.~\ref{fig4}, and the peak values are listed in Tab.~\ref{tab3}. We observe a clear hierarchy of the material platforms: the gold PAs support higher fields than the composite PAs, which are still higher than fields of the metallic \vo{} PAs. To characterize the nature of the hot spot (electric, magnetic, mixed), we define the electromagnetic contrast $C_\mathrm{EH}=(E-H)/(E+H)$, where $E$ and $H$ are the electric and magnetic field at the hot spot, respectively. The $C_\mathrm{EH}$ takes values between $-1$ (pure magnetic hot spot) and 1 (pure electric hot spot). The hot spots of the gold PAs are straight (the bowtie supports an electric hot spot and the diabolo supports a magnetic hot spot). A similar statement holds for the metallic \vo{} PAs, with the diabolo supporting a mixed hot spot. On the other hand, the composite PAs support mixed hot spots with a slight preference of the electric field ($C_\mathrm{EH}$ is near zero and positive), with only a minor difference between the \vowtie{} and the \diavolo{} PAs. Their functionality thus differs from the functionality of the homogeneous bowtie and diabolo PAs, with a specific field of applications.  

\begin{table}[h!]
\caption{\label{tab3}
Normalized electric ($E$) and magnetic ($H$) field and their contrast $C_\mathrm{EH}$ at the hot spot of the bowtie PAs (gold bowtie, metallic \vo{} bowtie, composite \vowtie{}) and the diabolo PAs (gold diabolo, metallic \vo{} diabolo, composite \diavolo{}).
}
\begin{tabular}{lccc||ccc}
\hline
\hline
& & bowtie & & & diabolo & \\
material & $E$ & $H$ & $C_\mathrm{EH}$ & $E$ & $H$ & $C_\mathrm{EH}$ \\  
\hline
gold & 13.8 & 2.40 & 0.703 & 6.32 & 26.7 & $-0.617$ \\
\vo & 5.84 & 1.25 & 0.647 & 2.45 & 2.48 & $-0.00620$ \\
composite & 9.86 & 6.10 & 0.235 & 6.82 & 5.25 & 0.130 \\ 
\hline
\hline
\end{tabular}
\end{table}

Finally, we inspect the switching performance of the PAs. We define several switching metrics (SMs): A shift of the LSP resonance $\Delta E=E_\mathrm{B}-E_\mathrm{D}$ ($E^\mathrm{B}$ and $E^\mathrm{D}$ are the LSP resonance energies of the bowtie and the diabolo, respectively), a change in the electromagnetic contrast $\Delta C_\mathrm{EH}=C_\mathrm{EH}^\mathrm{B}-C_\mathrm{EH}^\mathrm{D}$ (the superscripts B and D again refer to the bowtie and the diabolo PAs), and a change in the scattering-absorption contrast (SAC) $\Delta C_\mathrm{SA}=C_\mathrm{SA}^\mathrm{B}-C_\mathrm{SA}^\mathrm{D}$, where the SAC defined as $\Delta C_\mathrm{SA}=(Q_\mathrm{scat}-Q_\mathrm{abs})/(Q_\mathrm{scat}+Q_\mathrm{abs})$ represents a normalized analogy of SAR, and $Q_\mathrm{scat}$, $Q_\mathrm{abs}$ are the peak values of the normalized scattering and absorption cross sections. The values of the SMs are listed in Tab.~\ref{tab4} for the composite \vowtie{} and \diavolo{} PAs, and for reference also for gold PAs (which, naturally, cannot be dynamically switched, but can be replaced during the fabrication of a specimen). 

\begin{table}[h!]
\caption{\label{tab4}
Switching metrics: The shift of the LSP resonance energy $\Delta E$, the change in the electromagnetic contrast $\Delta C_\mathrm{EH}$, and the change in the scattering-absorption contrast $\Delta C_\mathrm{SA}$, defined as the parameter value for the bowtie minus the parameter value for the diabolo. 
}
\begin{tabular}{lccc}
\hline
\hline
material & $\Delta E$ (eV) & $\Delta C_\mathrm{EH}$ & $\Delta C_\mathrm{SA}$ \\  
\hline
composite & 0.470 & 0.105 & 0.970 \\
gold & 0.690 & 1.32 & 0.403 \\
\hline
\hline
\end{tabular}
\end{table}

A switching from the \vowtie{} to the \diavolo{} PA (or, in the case of gold, a replacement of the bowtie with the diabolo) is accompanied by the spectral red shift of the lowest LSP resonance to about half of the original value. For composite PAs, the hot spot preserves its mixed character, while for the gold PAs there is a clear transformation from the electric to the magnetic hot spot. The composite PAs also undergo a pronounced change from the balanced scattering and absorption towards pure absorption. A moderate change towards the larger absorption is also observed for gold PAs, but the scattering dominates the optical response for them both.


The unique properties of the composite PAs predestinate them for specific application fields, distinct from gold bowtie and diabolo PAs. Here we discuss three prospective applications: precise plasmon-enhanced spectroscopy, switchable antireflection coatings, and plasmonic optical shutters and choppers.

Plasmon-enhanced (PE) spectroscopy exploits PAs to enhance the interaction of the probing light with the examined specimen, and has multiple flavors, including plasmon-enhanced photoluminescence~\cite{Kinkhabwala2009}, PE Raman spectroscopy~\cite{Wang2020}, surface-enhanced infrared absorption also known by its acronym as SEIRA~\cite{https://doi.org/10.1002/adma.201704896}, or PE electron paramagnetic resonance~\cite{https://doi.org/10.1002/smtd.202100376}. The role of the PAs in PE spectroscopy is twofold. In addition to the desired enhanced light-matter coupling, often proportional to the field enhancement~\cite{PhysRevApplied.13.054045}, PAs also contribute with the undesired direct light scattering and absorption, hindering quantitative analysis of the acquired spectra~\cite{https://doi.org/10.1002/smtd.202100376}. The PAs suitable for precise PE spectroscopy should thus provide high field enhancement while having low optical cross sections, and the composite \diavolo{} PAs, and to some extent also the \vowtie{} PAs, fulfill these requirements well.

Antireflection coatings are an integral part of many optical elements, including lenses, optical sources and detectors, or solar cells~\cite{C4NR00087K}. Plasmonic antireflection coatings offer several advantages over traditional designs based on the interference, such as broadband operation and tolerance to oblique incidence~\cite{C4NR00087K,doi:10.1021/acsphotonics.7b00410}. The composite \diavolo{} PAs with their negligible scattering and relatively high peak absorption (approximately $3\times$ the physical cross section of the PA) are well-suited for such coatings. A large area density of PAs might be needed for optimum performance of the coating, and at the same time, the near-field interaction between the neighboring PAs might be detrimental to the coating performance. These requirements can be met by randomized vertical stacking of individual PAs (realized, e.g., by dispersing the PAs in a continuous medium). \ccs{The ARC is absorptive, limiting the applications to those related to the absorption of the light - detectors, energy harvestors. I suppose we do not need to discuss this.} Switching of SAR accompanied by the large absorption finds additional applications, e.g., as thermal emitters or bolometers~\cite{10.1063/1.4767646}.

Optical shutters and choppers transform continuous light waves into pulses and are involved in various approaches of optical spectroscopy. The shutters and choppers based on the switching between the absorptive \diavolo{} and the semi-scattering \vowtie{} would offer a high switching speed and exceptional stability as no mechanical movement is required. Here we again stress the possibility to trigger the switching by the electrical or optical biasing~\cite{Wu2011_VO2-electric,Lei2015_VO2-optic} on the subpicosecond scale~\cite{Wegkamp2015_VO2ultrafast}, and the possibility to stabilize both states of the switch in a range of temperatures due to broad hysteresis of MIT in \vo{}~\cite{horák2024efficientnanoscaleimagingsolidstate}.

\ccs{Material benefits of the design can be mentioned: Fast switching, which can be triggered by numerous stimuli; switching near the room temperature, which is further tunable by doping; broad and engineerable hysteresis, which allows stabilization of both phases under the same conditions.}


{\it Conclusion}
We have presented a composite material platform for plasmonic switches based on the lateral arrangement of gold and \vo{}. This platform enables full functional switching, and complements homogeneous plasmonic switches based on \vo{}, where only ON/OFF switching is possible (i.e., switching between the plasmonic antenna functionality and weak optical response of a dielectric). As the case study, we addressed the composite \vowtie{} and \diavolo{} PAs, formed by two metallic wings connected with the \vo{} bridge either in its insulating or conductive states, and compared them with their metallic counterparts (both gold and metallic \vo{}). The hot spots produced by the composite PAs exhibited mixed electric-magnetic character, which was only weakly altered upon switching. On the other hand, we observed a strong modification of the scattering-absorption contrast and the spectral position of the hot spots upon switching. Unique properties of the composite \vowtie{} and \diavolo{} PAs open the prospect for uncommon applications in precise plasmon-enhanced spectroscopy, switchable antireflection coatings, or optical shutters and choppers.

The data that support the findings of this article are openly available~\cite{data}.

\begin{acknowledgments}
We acknowledge support from the Ministry of Education, Youth, and Sports of the Czech Republic, projects No.~CZ.02.01.01/00/22\_008/0004572 (QM4ST) and LM2023051 (CzechNanoLab). RŘ was supported by Brno University of Technology (project No.~CEITEC VUT-J-25-8860). JK acknowledges support from the Brno Ph.D.~Talent Scholarship, funded by the Brno City Municipality. We acknowledge Petr Klenovský of Masaryk University in Brno for proposing the application as optical choppers.
\end{acknowledgments}

\del{
\begin{authorcontributions}
All authors have accepted responsibility for the entire content of this manuscript and approved its submission.
\end{authorcontributions}

\begin{conflictofinterest}
Authors state no conflict of interest.
\end{conflictofinterest}

\begin{dataavailabilitystatement}
The datasets generated during and/or analyzed during the current study are available in the Zenodo repository, doi:10.5281/zenodo.17552778.
\end{dataavailabilitystatement}
}

\bibliographystyle{apsrev4-2}

%

\end{document}